# Half a Century with QUARKS*

## (A superficial survey)

Vladimir A. Petrov

Institute for High Energy Physics

A brief account of interesting moments in the genesis of the quark paradigm is presented.

When preparing the program of this workshop the Organizing Committee couldn't miss such a subject as the $50^{th}$ anniversary of the quark model. Unfortunately we failed to have a professional historian of science so I was asked by the Organizing Committee to present (to the best of my limited abilities) at least a brief communication.

I figured that it would be quite impertinent to repeat ABC of SU(3) and its representations and so on in front of an audience of particle physicists.
It's also not my task to tell about successes of the quark model so well described in hundreds of excellent books. Instead I am to draw your attention to some moments which seem to me curious.

## Milestones

- (1963)1964 - the rise and propagation of the idea
- 1965 – colouring
- 1967-1973 – experimental "proof" that quarks exist
- 1973 – formulation of the strong interaction theory
  as QCD = $SU_{flavour}(3)$ x $SU_{colour}(3)$
- 1974 – first heavy quark
- 1977 – second heavy quark
- 1995 – third heavy quark (after a long and torturous search)
- QCD = $SU_{flavour}(6)$ x $SU_{colour}(3)$
- 2014 – still with 6 coloured quarks

## Some digression on literary modernism

In 1939 a novel "Finnegans Wake" appeared on the shelves of bookstores. The author, Irish poet and novelist James Joyce, was already well known as one of the prominent figures of the avant-garde modernism and one of the (if not the first) inventors of the "stream of consciousness" style.

---

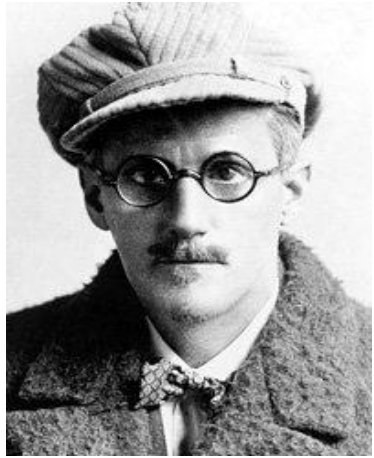

*James Joyce*

Long afterwards its publication the novel was qualified as *" one of the 100 best English-language novels of the 20th century".*
In order that you could feel the fairness of such a high praise let me give you a passage from the book:
*" The fall (bababadalgharaghtakamminarronnkonnbronntonner ronntuonnthunntrovarrhounawnskawntoohoohoordenenthurnuk!) of a once wallstrait oldparr is retaled early in bed and later on life down through all christian minstrelsy."*

Don't think I show this quotation for fun. Actually Joyce's book was and is a subject of a thorough study by many serious scholars worldwide. Here is the title of a thesis submitted in total fulfilment of the requirements for the degree of Doctor of Philosophy to the Australian Defence Force Academy, University of New South Wales:

"The Picture and the Letter: Male and Female Creativity in James Joyce's "Finnegans Wake""
                        by William L. Miller, 1996

However our interest is much narrower and relates to the following, now famous, lines:

    "*Good-bark, goodbye! Now follow we out by Starloe!*
    *— **Three quarks for Muster Mark!***
    *Sure he hasn't got much of a bark*
    *And sure any he has it's all beside the mark…*"

First of all what is "quark?" In vocabularies one can find synonyms:
"*curd*" or "*nonsense*"…And "*muster*'?
Here it is: "*a number of things or persons assembled on a particular occasion;*
*a collection, as a muster of peacocks*" ( Samuel Johnson. A Dictionary of the English Language, 1755.)
It seems that only the words "three" and "Mark" do not cause a confusion.

# Gell-Mann

Well, these inquiries can hardly explain us why on the earth a schoolboy in 1939 New York decided to read a strange book, evidently not for children ( abstruse, however, for an adult reader as well) full with strange things like e.g. "three quarks". The number "three" sure was at boy's everyday disposal and maybe when preparing his math homework he calculated the products:

$$3 \times 3 = 9,$$

$$3 \times 3 \times 3 = 27, ...$$

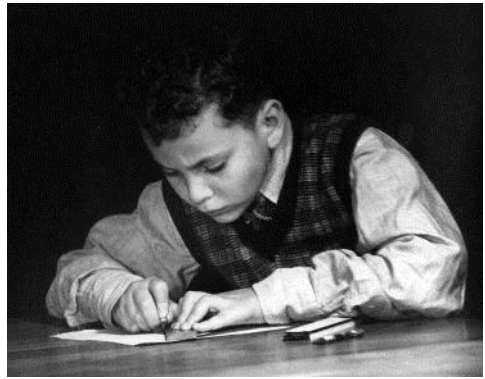

*Murray Gell-Mann, 10-year-old, New York, 1939*

But actually "three quarks" came to his mind much later when he was a young but already famous theorist in several branches of physics. Now, 25 years later, the products looked in a somewhat new way:

$$3 \otimes \overline{3} = 8 \oplus 1,$$

$$3 \otimes 3 \otimes 3 = 10 \oplus 8 \oplus 8 \oplus 1$$

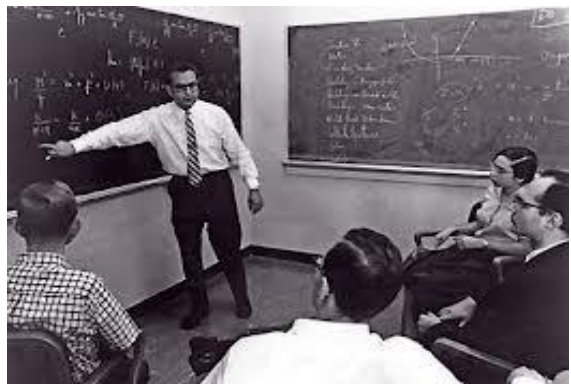

*Murray Gell-Mann, 25 years later, Caltech, Pasadena*

However the path to three quarks started from the number 8. As is well known by the beginning of 1960s some theorists trying to detect at least some order in an increasing flow of newly discovered strongly interacting particles mastered a powerful tool - the theory of Lie groups describing unitary symmetry. Gell-Mann, who was together with Yuval Ne'eman one of the most active proponents of this new trend in particle physics dubbed it "Eightfold Way" taking as an example the "Noble Eightfold Path", one of the principal teachings of Buddha, who described it as the way leading to the cessation of suffering and the achievement of self-awakening. Principles of this doctrine are shown below:

| |
| --- |
| *1. Right view* |
| *2. Right intention* |
| *3. Right speech* |
| *4. Right action* |
| *5. Right livelihood* |
| *6. Right effort* |
| *7. Right mindfulness* |
| *8. Right concentration* |

As we know, unitary symmetry based on SU(3) with eight generators ( thereof the "Eightfold Way") appeared extremely successful when describing the hadrons as members of unitary multiplets and , in fact, led particle physicists to " the cessation of suffering and the achievement of self-awakening". The awakening was very impressive with a flow of papers on symmetry approaches to the hadron and even lepton spectra.

Nonetheless, the existence of the fundamental representation of SU(3) with no particles ascribed to it definitely worried Gell-Mann and finally he dared to propose that there exist some seemingly fictitious "particles" which he called "quarks" (presumably known to him since his childhood) which can be combined to give the properties of all the known and predict new, still unknown, hadrons.

Volume 8, number 3      PHYSICS LETTERS      1 February

A SCHEMATIC MODEL OF BARYONS AND MESONS *

M. GELL-MANN

*California Institute of Technology, Pasadena, California*

Received 4 January 1964

Gell-Mann was very careful and didn't insist that quarks are real constituents of hadrons. Maybe thereof a modest adjective "schematic" in the title. One may suggest that was a kind of Gell -Mann's prevision of confinement.

## Zweig

Quite soon it turned out that Gell-Mann was not the only player in town. Young American (though born in Moscow in 1937) theorist George Zweig, that time at CERN, published a CERN preprint dated 17 January 1964 where a SU(3) model for hadron classification was suggested with 3 constituents called "aces".

Everybody knows that there are four aces in the deck of cards, not three. It seems that playing cards was not a forte of the young PhD. On the other hand couldn't it be a subconscious expectation of charm?

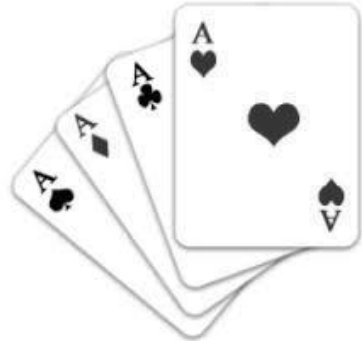

*Aces*

AN SU$_3$ MODEL FOR STRONG INTERACTION SYMMETRY AND ITS BREAKING

G. Zweig *)
CERN — Geneva

8182/TH. 401
17 January 1964

By no means was it a replica of the Gell-Mann paper. Zweig's work was quite lengthy (actually it was the first part of the whole paper with a detailed consideration of many physics consequences and predictions) in contrast to a concise Gell - Mann's paper. Some bizarre story happened to Zweig's work: a renowned Belgian theorist Léon Van Hove who was the responsible person at CERN TH not only prevented submission of the paper for publication but even insisted to cancel already announced seminar! Though afterwards Van Hove cited Zweig as one of the authors of the quark model. Certainly this episode didn't and couldn't spoil the value of fairly original Zweig's scheme and its application to hadrons proved to be very successful.

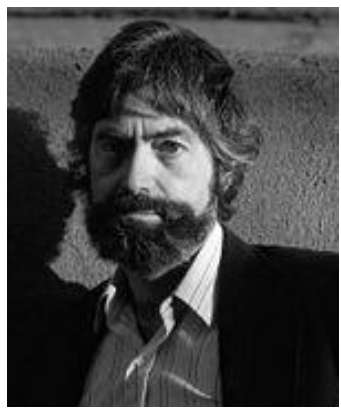

*George Zweig*

## Long hesitation

It is interesting to note that both authors – Gell-Mann and Zweig - have mentioned elsewhere later that the idea of fundamental entities came to them long before, back to the spring of 1963. However during the subsequent period both didn't show a smallest trace of it in theirs publications. Thus, Gell-Mann only a few days before submitting his paper on quarks submitted a paper where the old "eightfold way" was diligently applied to nonleptonic weak decays and where there was no even a hint of something like quarks.

VOLUME 12, NUMBER 6       PHYSICAL REVIEW LETTERS       10 FEBRUARY 1964

NONLEPTONIC WEAK DECAYS AND THE EIGHTFOLD WAY*

Murray Gell-Mann
California Institute of Technology, Pasadena, California
(Received 2 January 1964)

As to Zweig, after finishing his PhD thesis ( formally defended on 1 January 1964 but actually made almost a year before ) entitled
"Two topics in elementary particle physics: The reaction $\gamma n \to \pi N$ at high energies. K leptonic decay and partially conserved currents"
he only produced (in August 1963) a voluminous Caltech preprint again devoted to the high-energy pion photoproduction on nucleons and which didn't bear anything close to something like the ace idea.

How to explain such a hesitation? Leaving aside the funny resemblance of such a nine-month period to the usual time for carrying the fetus, I think that probably adding three more and unknown entities to a plenty of new known particles (hadrons) seemed not very ingenious. It's also worth to notice that filling the fundamental triplet with existing particles (proton, neutron and lambda-baryon) was already tried by Japanese theorist Sakata but afterwards this model failed. One another reason of this delay was maybe the fact that when ( March 1963) Robert Serber told to Gell-Mann about his reflections on a possibility to use fundamental triplets and anti-triplets for construction of baryons and mesons Gell-Mann immediately realized that such fundamental particles should have fractional electric charges, quite unpleasant feature as it could seem that time.

## One plus two

Recently I've learned from a short note by Alvaro de Rújula published in "CERN Courrier" that there was one more physicist who independently came to the idea of the fundamental triplet particles. It was a Swiss theorist André Petermann who worked at the Theory Division of CERN. It is of interest to notice that George Zweig was in this same division as a post-doctoral fellow in 1963-64. The title of Petermann's article, in contrast to those by Gell-Mann and Zweig, looks quite modest and even misleading:
" Properties of Strangeness and Mass Formula for Vector Mesons" (see below, in French)

# PROPRIÉTÉS DE L'ÉTRANGETÉ ET UNE FORMULE DE MASSE POUR LES MÉSONS VECTORIELS

A. PETERMANN

*CERN, Genéve* 23

**Abstract:** A mass-formula for vector mesons is proposed, and the role of strangeness in mass-formulae discussed.

The abstract also doesn't tell more than the title. However, if you go further and read the text you'll see immediately that the author pushes forward neither more nor less a new model of the hadron structure. You easily realize that "quarks" or "aces" have their counterparts in Petermann's "elementary spinor particles" with all attributes including fractional charges.

## Obscurity

There are no references to Gell-Mann and Zweig though Petermann's article was published very late, in March 1965, and the author could revise the paper to include congenial colleagues into the reference list.
However, neither Gell-Mann nor Zweig never mentioned Petermann as well.
Quite a confusing situation: three different authors who could fairly cross each other ( Gell-Mann was for a while Zweig's supervisor, Petermann and Zweig worked in the CERN Theory Division in 1963-64) wrote three independent articles concerning *grosso modo* the same subject within some 20 days. Two of them were published in well-known journals "Physics Letters" and "Nuclear Physics" and one as a CERN preprint.  CERN preprints were sent and read worldwide by relevant theorists.
While two of these three authors were rendered all due honors (including the Nobel Prize to Gell-Mann), the third one was put in oblivion. Maybe it was the French language of Petermann's article which appeared an insurmountable obstacle? Seems highly unlikely. E.g. the fact that a paper by Petermann and Stueckelberg, published in 1953 in French, did not prevent the international community of physicists to recognize it as a milestone in the development of the famous renormalization group method with multiple citations. Up to now I couldn't find any relevant reference to Petermann's paper in question.
Sure, one might suspect that Petermann could plagiarize and revise his paper taking use of the Gell-Mann and Zweig papers (thereof so long publication time?) but perennial colleagues of André Petermann (who unfortunately passed away in 2011) do not admit even a shadow of doubt in his high honesty. I have also to add that there is no editor's note of a revision of the Petermann article.
This intriguing story definitely has some dead ends and maybe one day they will be clarified.
At the moment I limit myself with one more quotation from James Joyce:

*"Thus the unfacts, did we possess them, are too imprecisely few to warrant our certitude..."*
(James Joyce, *Finnegans Wake* )

**Conclusion**

With all that I believe it would be fair to agree that three quarks (under various names) were discovered independently by the "Splendid Triplet": Petermann, Gell-Mann and Zweig.

Let us look at heroes of this talk and give them all the due tribute with our burst of applause!

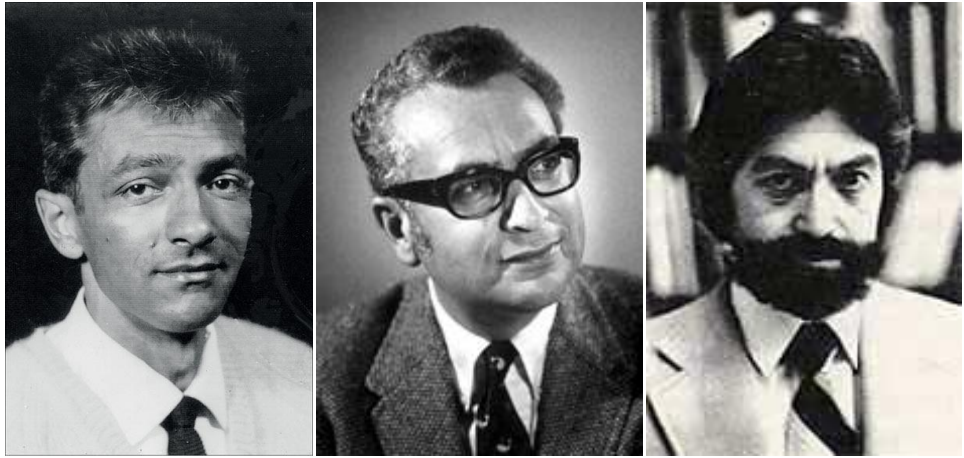

*André Petermann*  *Murray Gell-Mann*  *George Zweig*